\documentclass[showpacs,preprintnumbers]{revtex4}
\usepackage{graphicx}
\usepackage{color}

\begin{document}

\date{\today}
\title{Graphene nanoribbon based spaser}
\author{Oleg L. Berman$^{1,2}$, Roman Ya. Kezerashvili$^{1,2}$, and Yurii E.
Lozovik$^{3,4}$}
\affiliation{\mbox{$^{1}$Physics Department, New
York City College of Technology, The
City University of New York,} \\
Brooklyn, NY 11201, USA \\
\mbox{$^{2}$The Graduate School and University Center, The
City University of New York,} \\
New York, NY 10016, USA \\
\mbox{$^{3}$Institute of Spectroscopy, Russian Academy of
Sciences,} \\
142190 Troitsk, Moscow Region, Russia \\
\mbox{$^{4}$Moscow
Institute of Physics and Technology (State University), 141700,
Dolgoprudny, Russia }}

\begin{abstract}

A novel type of spaser with the net amplification of surface
plasmons (SPs) in doped graphene nanoribbon is proposed. The
plasmons in THz region can be generated in a dopped graphene
nanoribbon due to nonradiative excitation by emitters like two level
quantum dots located along a graphene nanoribbon.  The minimal
population inversion per unit area, needed for the net amplification
of SPs in a doped graphene nanoribbon is obtained. The dependence of
the minimal population inversion on the surface plasmon wavevector,
graphene nanoribbon width, doping and damping parameters necessary
for the amplification of surface plasmons in the armchair graphene
nanoribbon is studied.

\vspace{0.1cm} 

\end{abstract}

\maketitle




{}



\section{Introduction}

\label{intro}

The essential achievements in nanoscience and nanotechnology during the past
decade lead to great interest in studying nanoscale optical fields. The
phenomenon of surface plasmon amplification by stimulated emission of
radiation (spaser) was proposed in Ref.~\onlinecite{Bergman_Stockman} (see
also Refs.~\onlinecite{Stockman_review,Protsenko}). Spaser generates
coherent high-intensity fields of selected surface plasmon (SP) modes that
can be strongly localized on the nanoscale. The properties of localized
plasmons are reviewed in Refs.~%
\onlinecite{Agranovich,Klyuchnik,Zayats,Bozhevolnyi}. The spaser consists of
an active medium formed by two-level systems (semiconductor quantum dots
(QDs) or organic molecules) and a plasmon resonant nanosystem where the
surface plasmons are excited. The emitters transfer their excitation energy
by radiationless transitions through near fields to a resonant plasmon
nanosystem.

By today theoretical and experimental studies are focused on
metal-based spasers, where surface plasmons are excited in different
metallic nanostructures of different geometric shapes. A spaser
consisted of the nanosystem formed by the V-shaped silver
nanoinclusion embedded in a dielectric host with the embedded PbS
and PbSe QDs was considered ~in Ref. \cite{Bergman_Stockman}.
  A spaser formed by a silver spherical nanoshell on a
dielectric core with a radius of $10-20\ \mathrm{nm}$, and
surrounded by two dense monolayers of nanocrystal QDs  was
considered in Ref.~\onlinecite{Stockman}. The SPs propagating along
the bottom of a groove (channel) in the metal surface were studied
in Ref.~\onlinecite{Lisyansky}. The SPs are assumed to be coherently
excited by a linear chain of QDs at the bottom of the channel. It
was shown that for realistic values of the system parameters, gain
can exceed loss and plasmonic lasing in
 a ring or linear channels in the silver surface
surrounded by a linear chain of CdSe QDs can occur. In Refs.
\cite{Andrianov_ol,Andrianov_oe,Andrianov_prb1,Andrianov_prb2} the
spaser formed by the metal sphere surrounded by the two-level
quantum dot was studied theoretically. The spaser consisting of the
spherical gain core, containing two-level systems, coated with a
metal spherical plasmonic shell was theoretically analyzed in Ref.
\cite{Baranov}. The experimental study of the spaser formed by $44
 \ \mathrm{nm}$ diameter nanoparticles with the gold spherical core surrounded
by dye-doped silica shell was performed in
Refs.~\cite{Noginov_2007,Noginov}. In this experiment the emitters
were formed by dye-doped silica shell instead of QDs. It was
demonstrated that a two-dimensional array of a certain class of
plasmonic resonators supporting coherent current excitations with
high quality factor can act as a planar source of spatially and
temporally coherent radiation~\cite{Zheludev}. This structure
consists of a gain medium slab supporting a regular array of
silver asymmetric split-ring resonators. The spaser formed by $55\ \mathrm{nm%
}$-thick gold film with the nano-slits located on the silica
substrate surrounded by PbS QDs was experimentally studied in
Ref.~\onlinecite{Plum}. Room temperature spasing of surface plasmon
polaritons at $1.46\ \mathrm{\mu m}$ wavelength has been
demonstrated by sandwiching a gold-film plasmonic
waveguide between optically pumped InGaAs quantum-well gain media~\cite%
{Flynn}.

Since plasmons can be excited also in graphene, and damping in
graphene is much less than in
metals~~\cite{Varlamov,Falkovsky_prb,Falkovsky_conf}, we propose to
use graphene nanoribbon surrounded by semiconductor QDs as the
nanosystem for the spaser. Plasmons in graphene provide a suitable
alternative to plasmons in noble metals, because they exhibit much
tighter confinement and relatively long propagation distances, with
the advantage of being highly tunable via electrostatic
gating~\cite{Koppens}. Besides, the graphene-based spaser can work
in THz frequency regime. Recently there were many experimental and
theoretical studies devoted to graphene known by
unusual properties in its band structure~\cite{Castro_Neto_rmp,Das_Sarma_rmp}%
. The properties of plasmons in graphene were discussed in Refs.~\cite%
{Hwang_Das_Sarma,Lozovik_u,Mikhailov,Geim_plasmons}. The electronic
properties of graphene nanoribbons depend strongly on their size and
geometry~\cite{Brey_Fertig_01,Brey_Fertig}. The frequency spectrum
of oblique terahertz plasmons in graphene nanoribbon arrays was obtained~\cite%
{Popov}. Besides, graphene-based spaser seems to meet the new
technological needs, since it works at the infrared  (IR)
frequencies, while the metal-based spaser works at the higher
frequencies. Let us mention that the graphene-based photonic two-
and one-dimensional crystals proposed in Refs.~\onlinecite{BBKKL,BK}
also can be used effectively as the frequency filters and waveguides
for the far infrared region of electromagnetic spectrum.

In this Paper we propose the graphene nanoribbon based spaser
consisting of a graphene nanoribbon surrounded by semiconductor QDs.
The QDs excited by the laser pumping nonradiatively transfer their
excitation to the SPs localized at the graphene nanoribbon, which
results in an increase of intensity of the SP field. We calculate
the minimal population inversion that is the difference between the
surface densities of QDs in the excited and ground states needed for
the net SP amplification and study its dependence on the surface
plasmon wavevector, graphene nanoribbon width  at fixed temperature
for different doping and damping parameters for the armchair
graphene nanoribbon.

The paper is organized in the following way. In Sec.~\ref{dev} the minimal
population inversion for the graphene-based spaser is obtained. The
discussion of the results and conclusions follow in Sec.~\ref{disc}.

\section{Surface plasmon amplification}

\label{dev}

The system under consideration is the graphene nanoribbon, which is
the stripe of graphene at $z=0$ in the plane $(x,y)$, that is
infinite in $x$ direction and has the width $W$ in $y$ direction.
This stripe is surrounded by the deposited dense manolayers of
nanocrystal quantum dots with the dielectric constant $\varepsilon
_{d}$ at $z<0$ and $z>0$. When the quantum dots are optically
pumped, the resonant nonradiative transmission occurs by creating a
surface plasmon localized in the graphene nanoribbon. Our goal is to
show that amplification by QDs can exceed absorption of the surface
plasmon in the graphene nanoribbon. As a result we obtain an
increase of intensity of the surface plasmon field. In other words,
the competition between gain and loss of the surface plasmon field
in the graphene nanoribbon will result in favor of the gain.

Below we derive the expression for the minimal population inversion
per unit area $N_{c}$, needed for the net amplification of SPs in a
doped graphene nanoribbon from the condition that for the regime of
the plasmon amplification the rate $\partial \bar{U}/\partial t$ of
the transfer of the average energy of the QDs is greater than the
heat released per unit time $\partial Q/\partial t$ due to the
absorption of the energy of the plasmon field in the graphene
nanoribbon.

 Let us start from the Poynting theorem for the rate of the
transfer of the energy density from a region of space $\partial \mathcal{W}%
/\partial t=-\mathrm{div}\vec{S}$, where $\vec{S}$ is the Poynting
vector and assume that the plasmon frequency equals the QD
transition frequency. From the other side the rate of the
transferred energy related to the rates of the average energy of the
QDs and the heat released due to the absorption of the energy by the
graphene nanoribbon can be presented as
\begin{eqnarray}
\label{wuq}
-\frac{\partial }{\partial t}\int \mathcal{W}dV=\frac{\partial \bar{U}}{%
\partial t}-\frac{\partial Q}{\partial t}\ ,
\end{eqnarray}
where  $V$ is the volume of the system. Therefore, from the Poynting
theorem  we have the following expression
\begin{eqnarray}
\label{S01}
\int \nabla \cdot \vec{S}dV=\frac{\partial \bar{U}}{\partial t}-\frac{%
\partial Q}{\partial t}\ \ .
\end{eqnarray}

Let us consider now each term in the left  hand side
 of Eq.~(\ref{S01}) separately. The excitation
causing  the generation of plasmons in the graphene nanoribbon comes
from the transitions in the QDs between the excited and ground
states. The average
energy $\bar{U}$ of the QDs characterized by the dipole moment is given by~%
\cite{Tamm}
\begin{eqnarray}
\label{uav1}  \bar{U}=\frac{1}{2}\int \vec{P}\cdot \vec{E}dV\ ,
\end{eqnarray}
where $\vec{E}$ is the electric field of the graphene nanoribbon
plasmon, and $\vec{P}$ is the polarization of QDs, which is the
average total dipole moment of the unit of the volume $V$. When the
plasmon frequency $\omega $ equals the QD transition frequency, and,
$\vec{E}\sim \exp (-i\omega t)$ and $\vec{P}\sim \exp (-i\omega t)$,
the relation between the polarization of QDs $\vec{P}$ and electric
field of the graphene nanoribbon plasmon $\vec{E}$ has the
form~\cite{Lisyansky}
\begin{eqnarray}
\label{Lis} \vec{P}=-ik\frac{\tau _{p}|\mu |^{2}n_{0}}{\hbar
}\vec{E}\ ,
\end{eqnarray}
where $k=9\times 10^{9}\ \mathrm{N\times m^{2}/C^{2}}$, $n_{0}$ is
the difference between the concentrations of quantum dots in the
excited and ground states, $\tau _{p}$ is the inverse line width,
and $\mu $ is the average off-diagonal element of the dipole moment
of a single QD.

Substituting Eq.~(\ref{Lis}) into Eq.~(\ref{uav1}), we obtain the rate of
the transfer of the average energy of the QDs
\begin{eqnarray}  \label{uav11}
\frac{\partial \bar{U}}{\partial t} = \int \omega \mathrm{Im} \left(\vec{E}%
\cdot\vec{P}^{\ast}\right) d V = \omega k\frac{\tau _{p}|\mu |^{2}}{\hbar }%
\int n_{0} |\vec{E}|^{2} d V \ .
\end{eqnarray}

We assume that the distances between the quantum dots are small, so
their effect on a plasmon is the same as that of a continuous
(constant) gain distribution along the graphene nanoribbon.  We
consider the two-dimensional graphene nanoribbon at $z=0$ and assume
it is infinite in $x$ direction, has the width $W$ in $y$ direction
and therefore, $n_{0}=N_{0}\eta (y,-W/2,W/2)\delta (z)$, where
$N_{0}$ is the difference between the numbers of the excited and
ground state quantum dots per unit area of the graphene nanoribbon,
and $\eta (y,-W/2,W/2)=1$ at $-W/2\leq y\leq W/2$, $\eta
(y,-W/2,W/2)=0$ at $y<-W/2$ and $y>W/2$. Then, taking into account
 mentioned above, we obtain from Eq.~(\ref{uav11})
\begin{eqnarray}\label{dudt}
\frac{\partial \bar{U}}{\partial t} &=&\omega k\frac{\tau _{p}|\mu |^{2}}{%
\hbar }\int_{-\infty }^{+\infty }dx\int_{-\infty }^{+\infty }dy\int_{-\infty
}^{+\infty }dzN_{0}\eta (y,-W/2,W/2)\delta (z)|\vec{E}(x,y,z)|^{2}=
\nonumber  \label{uav111} \\
&=&\omega k\frac{\tau _{p}|\mu |^{2}N_{0}}{\hbar }\int_{-W/2}^{+W/2}dy%
\int_{-\infty }^{+\infty }dx|\vec{E}(x,y,0)|^{2}\ .
\end{eqnarray}%

Taking into account the spatial dispersion of the dielectric
function in the graphene
nanoribbon~\cite{Brey_Fertig_01,Brey_Fertig}, we use the following
expression for the rate of the heat  $\partial Q/\partial t$
released due to the absorption of the energy of the plasmon field in
the graphene nanoribbon~\cite{Agranovich,Landau}
\begin{eqnarray}  \label{uav2}
\frac{\partial Q}{\partial t} &=& \int \omega \mathrm{Im} \varepsilon(%
\omega,q_{x})\eta(y,-W/2,W/2)|\vec{E}|^{2} d V  \nonumber \\
&=& \omega \mathrm{Im} \varepsilon(\omega,q_{x}) \int_{-\infty}^{+\infty} d
x \int_{-W/2}^{+W/2} d y \int_{-\infty}^{+\infty} d z |\vec{E}(x,y,z)|^{2} \
.
\end{eqnarray}
where $\mathrm{Im} \varepsilon(\omega,q_{x})$ is the imaginary part of the
dielectric function $\varepsilon(\omega,q_{x})$ of graphene nanoribbon given
by Eq.~(\ref{epsilon}).

The plasmons in a graphene nanoribbon are excited due to the
radiation caused by the transitions from the excited state to the
ground state on the QDs. Therefore, according to the conservation of
energy, the regime of the amplification of the plasmons in the
graphene nanoribbon is established, if the rate of the transfer of
the average energy  $\partial \bar{U}/\partial t$  of the QDs
 given by Eq.~(\ref{dudt}) is greater than the heat
released rate $\partial Q/\partial t$ due to the absorption of the
energy of the plasmon field in the graphene nanoribbon:
\begin{eqnarray}
\label{ineq} \frac{\partial \bar{U}}{\partial t}>\frac{\partial
Q}{\partial t}\ .
\end{eqnarray}
Substituting Eqs.~(\ref{uav111}) and~(\ref{uav2}) into Eq.~(\ref{ineq}), we
get
\begin{eqnarray}
\label{comb}\omega k\frac{\tau _{p}|\mu |^{2}N_{0}}{\hbar }\int_{-W/2}^{+W/2}dy\int_{-%
\infty }^{+\infty }dx|\vec{E}(x,y,0)|^{2}>\omega \mathrm{Im}\varepsilon
(\omega ,q_{x})\int_{-\infty }^{+\infty }dx\int_{-W/2}^{+W/2}dy\int_{-\infty
}^{+\infty }dz|\vec{E}(x,y,z)|^{2}\ .
\end{eqnarray}
From Eq.~(\ref{comb}), one can obtain the condition for the
difference between the surface densities of the quantum dots in the
excited and ground state corresponding to the amplification of
plasmons:
\begin{eqnarray}
\label{N0} N_{0}>N_{c}=\frac{\mathrm{Im}\varepsilon (\omega
,q_{x})\int_{-\infty
}^{+\infty }dx\int_{-W/2}^{+W/2}dy\int_{-\infty }^{+\infty }dz|\vec{E}%
(x,y,z)|^{2}}{k\frac{\tau _{p}|\mu |^{2}}{\hbar }\int_{-\infty
}^{+\infty }dx\int_{-W/2}^{+W/2}dy|\vec{E}(x,y,0)|^{2}}\ ,
\end{eqnarray}
where $N_{c}$ is the critical density of the QDs required for the
amplification of the plasmons. The evaluation of the integrals in
Eq.~(\ref{N0}) requires the knowledge of the electric field of a
plasmon in a graphene nanoribbon. The electric field of a plasmon in
a graphene nanoribbon is derived in Appendix~\ref{ap.el}. Using
Eq.~(\ref{E}) for the electric field of a plasmon, we have:
\begin{eqnarray}  \label{E0}
|\vec{E}(x,y,0)|^{2}=E_{0}^{2}\left( 2q_{x}^{2}\cos
^{2}(q_{y}y)+q_{y}^{2}\right) \ ,
\end{eqnarray}
\begin{eqnarray}  \label{Exyz}
|\vec{E}(x,y,z)|^{2}=E_{0}^{2}e^{-2\alpha |z|}\left( 2q_{x}^{2}\cos
^{2}(q_{y}y)+q_{y}^{2}\right) \ ,
\end{eqnarray}
where $\alpha =\sqrt{q_{x}^{2}+q_{y}^{2}}$ and for the armchair nanoribbon
we have $q_{yn}=2\pi /(3a_{0})\left( (2M+1+n)/(2M+1)\right) $ at the width $%
W=(3M+1)a_{0}$ \cite{Brey_Fertig_01}, where $a_{0}$ is the graphene
lattice constant, $M$ is the integer. We will use $n=1$.
Substituting Eqs.~(\ref{E0}) and~(\ref{Exyz}) into Eq.~(\ref{N0}),
we obtain
\begin{eqnarray}
\label{N01} N_{0}>N_{c}=\frac{2\hbar \mathrm{Im}\varepsilon (\omega
,q_{x})\int_{-\infty }^{+\infty
}dx\int_{-W/2}^{+W/2}dy\int_{0}^{+\infty }dze^{-2\alpha z}\left(
2q_{x}^{2}\cos ^{2}(q_{y}y)+q_{y}^{2}\right) }{k\tau _{p}|\mu
|^{2}\int_{-\infty }^{+\infty }dx\int_{-W/2}^{+W/2}dy\left(
2q_{x}^{2}\cos ^{2}(q_{y}y)+q_{y}^{2}\right) } \ .
\end{eqnarray}
 Finally from Eq.~(\ref{N01}) we obtain
\begin{eqnarray}
\label{N02} N_{0}>N_{c}=\frac{\hbar \mathrm{Im}\varepsilon (\omega
,q_{x})}{\alpha k\tau _{p}|\mu |^{2}}\ .
\end{eqnarray}
Using Eqs.~(\ref{epsilon}) and (\ref{pi}) one can find $\mathrm{Im}%
\varepsilon (q_{x},\omega )$:
\begin{eqnarray}  \label{ime}
\mathrm{Im}\varepsilon (q_{x},\omega
)=-\frac{V_{00}(q_{x})f_{1}(q_{x},\beta ,\mu )g_{s}v_{F}q_{x}\omega
\gamma }{\pi \hbar \left( \left( \omega
^{2}-v_{F}^{2}q_{x}^{2}\right) ^{2}+\omega ^{2}\gamma ^{2}\right) }
\ ,
\end{eqnarray}
where $v_{F}$ is the Fermi velocity of electrons in graphene. The
plasmon frequency $\omega$ can be obtained at $\gamma = 0$ from the
condition $\mathrm{Re} \varepsilon (q_{x},\omega) = 0$ using Eqs.~(\ref%
{epsilon}) and~(\ref{pi}):
\begin{eqnarray}  \label{pl}
\omega^{2} = v_{F}^{2}q_{x}^{2} - \frac{V_{0,0}(q_{x})f_{1}(q_{x},\beta,%
\mu)g_{s}v_{F}q_{x}}{\pi\hbar} \ .
\end{eqnarray}

To perform the calculations, one should calculate the critical density $N_{c}
$ using Eq.~(\ref{N02}). $N_{c}$ is a function of the wave vector $q_{x}$,
the graphene nanoribbon width $W$, temperature $T$, and electron
concentration $n_{0}$ determined by the doping.

\section{Results and discussion}

\label{disc}

For our calculations we use the following parameters for the PbS and PbSe
QDs. Since the typical energy corresponding to the transition between the
ground and excited electron states for PbS and PbSe QDs synthesized with the
radii from $1 \ \mathrm{nm}$ to $8 \ \mathrm{nm}$ can be $0.7 \ \mathrm{eV}$%
, we use $\tau_{p} \approx 5.9 \ \mathrm{fs}$, and $|\mu| = 1.9 \times
10^{-17} \ \mathrm{esu} = 19 \ \mathrm{Debye}$ ($1 \ \mathrm{Debye} =
10^{-18} \ \mathrm{esu}$, $1 \ \mathrm{Debye} = 3.33564 \times 10^{-30} \
\mathrm{C\cdot m}$)~\cite{Stockman_QD}. Let us mention that the typical
frequency corresponding to the transition between the ground and excited
electron states for PbS and PbSe QDs, which is $f \approx 170 \ \mathrm{THz%
}$, matches the resonance with the plasmon frequency in the armchair
graphene nanoribbon~\cite{Brey_Fertig}. Therefore, PbS and PbSe QDs can be
used for the spaser considered here. The damping in graphene $\gamma =
\tau^{-1}$ determined by $\tau$ is assumed to be either $\tau = 1 \ \mathrm{%
ps}$ or $\tau = 10 \ \mathrm{ps}$ or $\tau = 20 \ \mathrm{ps}$~\cite%
{Neugebauer,Orlita,Dubinov,Emani}.

\begin{figure}[tbp]
\includegraphics[width=10cm]{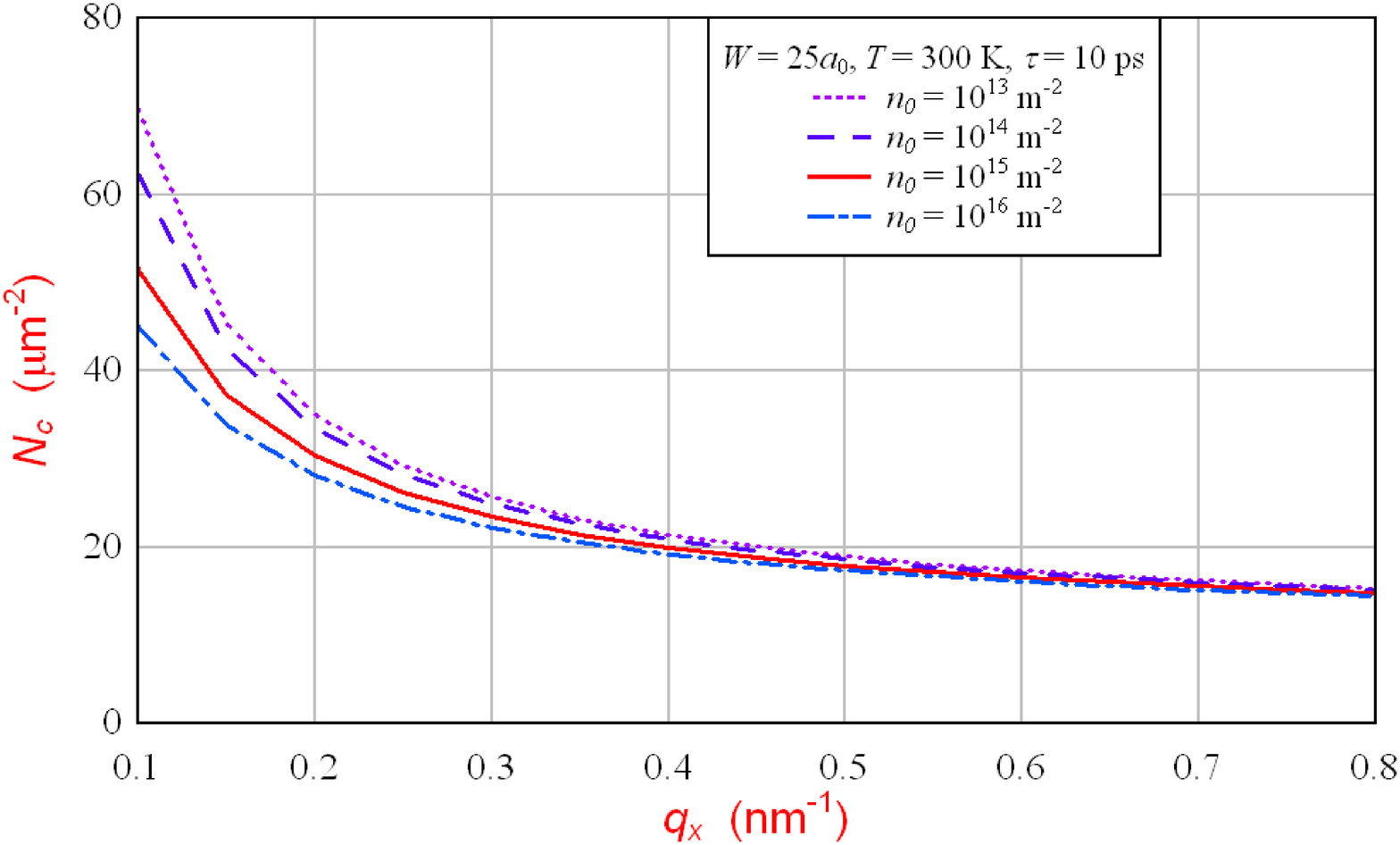}
\caption{The dependence of the critical density of the QDs $N_{c}$
required for the amplification of the signal on the wave vector
$q_{x}$ for the different doping electron densities $n_{0}$ at the
fixed width of the nanoribbon $W$, temperature $T$ and dissipation
time $\protect\tau$ corresponding to the damping.} \label{Fig1}
\end{figure}

\begin{figure}[tbp]
\includegraphics[width=10cm]{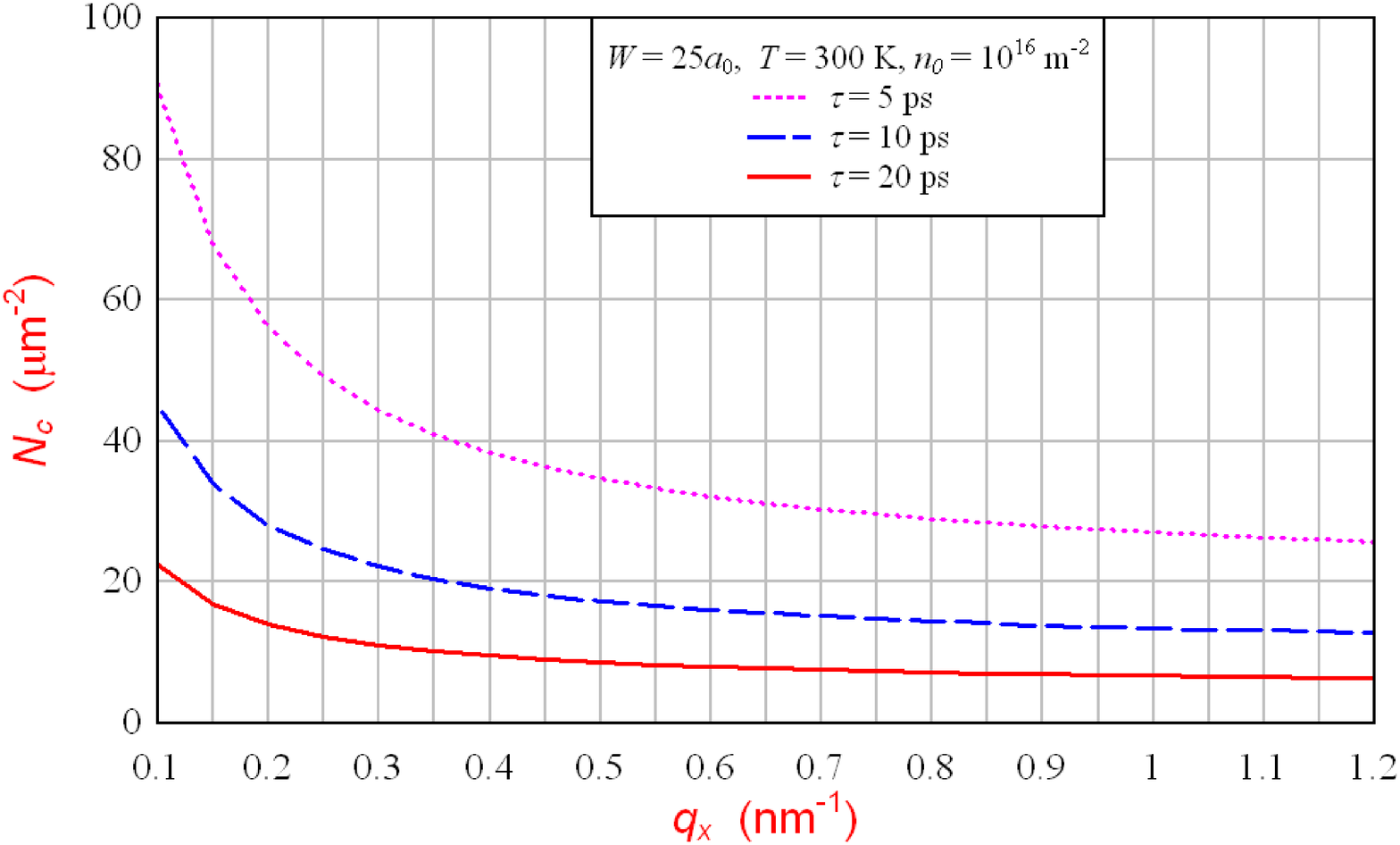}
\caption{The dependence of the critical density of the QDs $N_{c}$
required for the amplification of the signal on the wave vector
$q_{x}$ for the different dissipation time $\protect\tau$
corresponding to the damping at the fixed width of the nanoribbon
$W$, temperature $T$ and doping electron densitiy $n_{0}$.}
\label{Fig2}
\end{figure}

\begin{figure}[tbp]
\includegraphics[width=10cm]{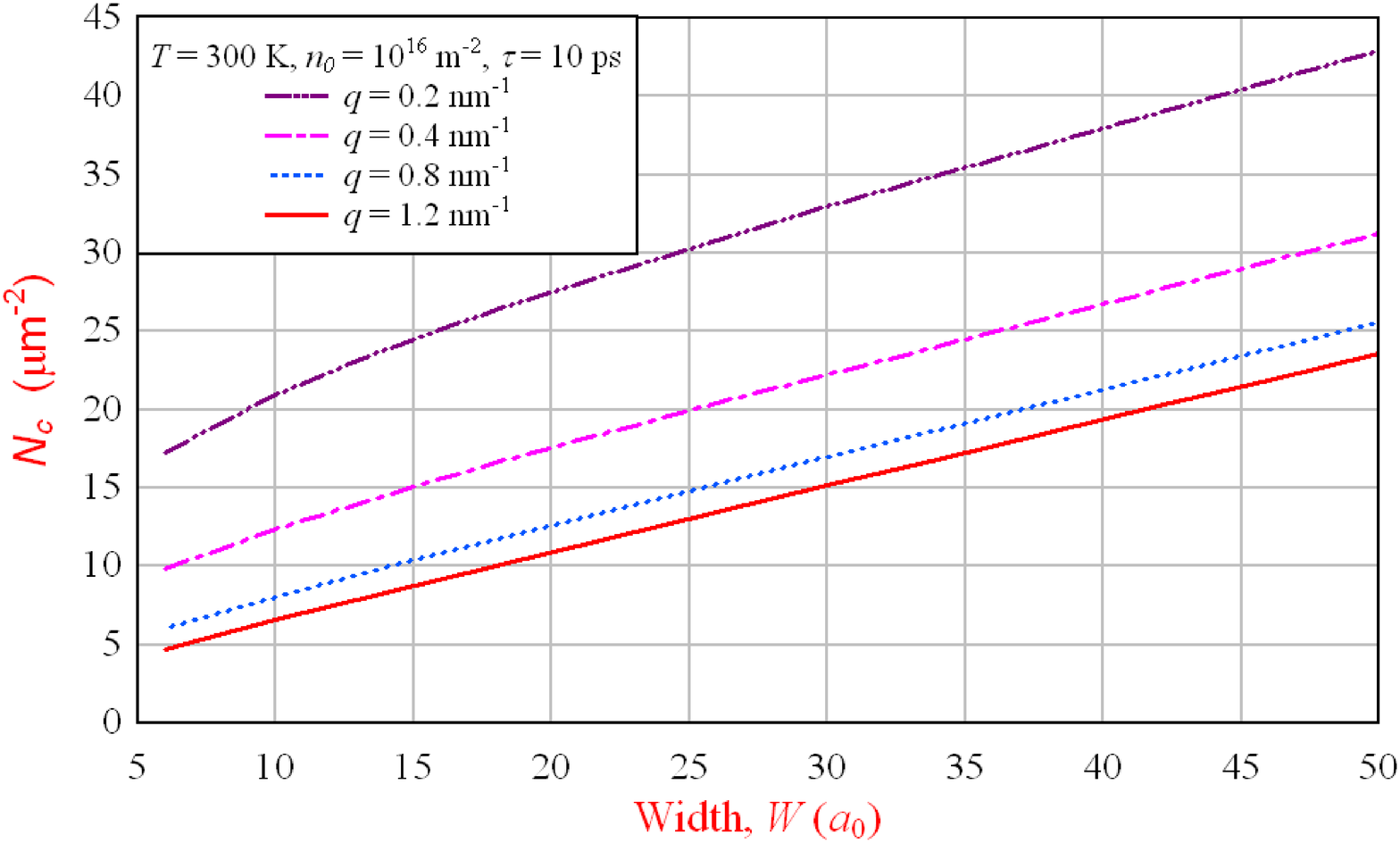}
\caption{The dependence of the critical density of the QDs $N_{c}$
required for the amplification of the signal  on the width of the
nanoribbon $W$ at
the different wave vector $q_{x}$ for the fixed dissipation time $\protect%
\tau$ corresponding to the damping, temperature $T$ and doping electron
densities $n_{0}$.}
\label{Fig3}
\end{figure}

\begin{figure}[tbp]
\includegraphics[width=10cm]{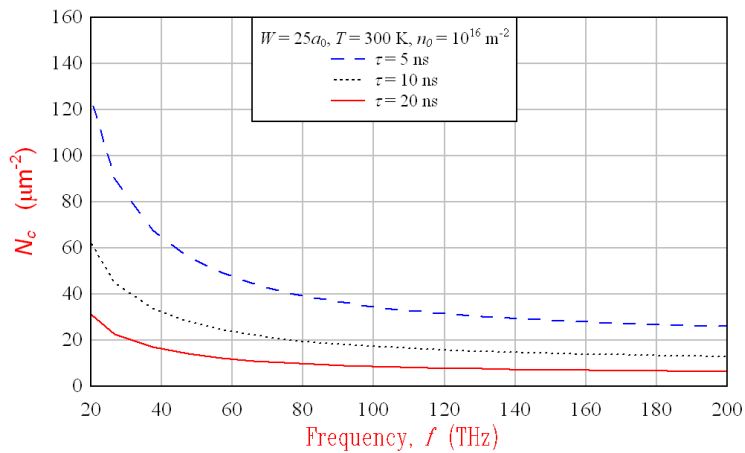}
\caption{The dependence of the critical density of the QDs $N_{c}$
required for the amplification of the signal  on the frequency $f$
at the different dissipation time $\protect\tau$ corresponding to
the damping for the fixed width $W$, temperature $T$ and doping
electron density $n_{0}$.} \label{Fig4}
\end{figure}

\begin{figure}[tbp]
\includegraphics[width=10cm]{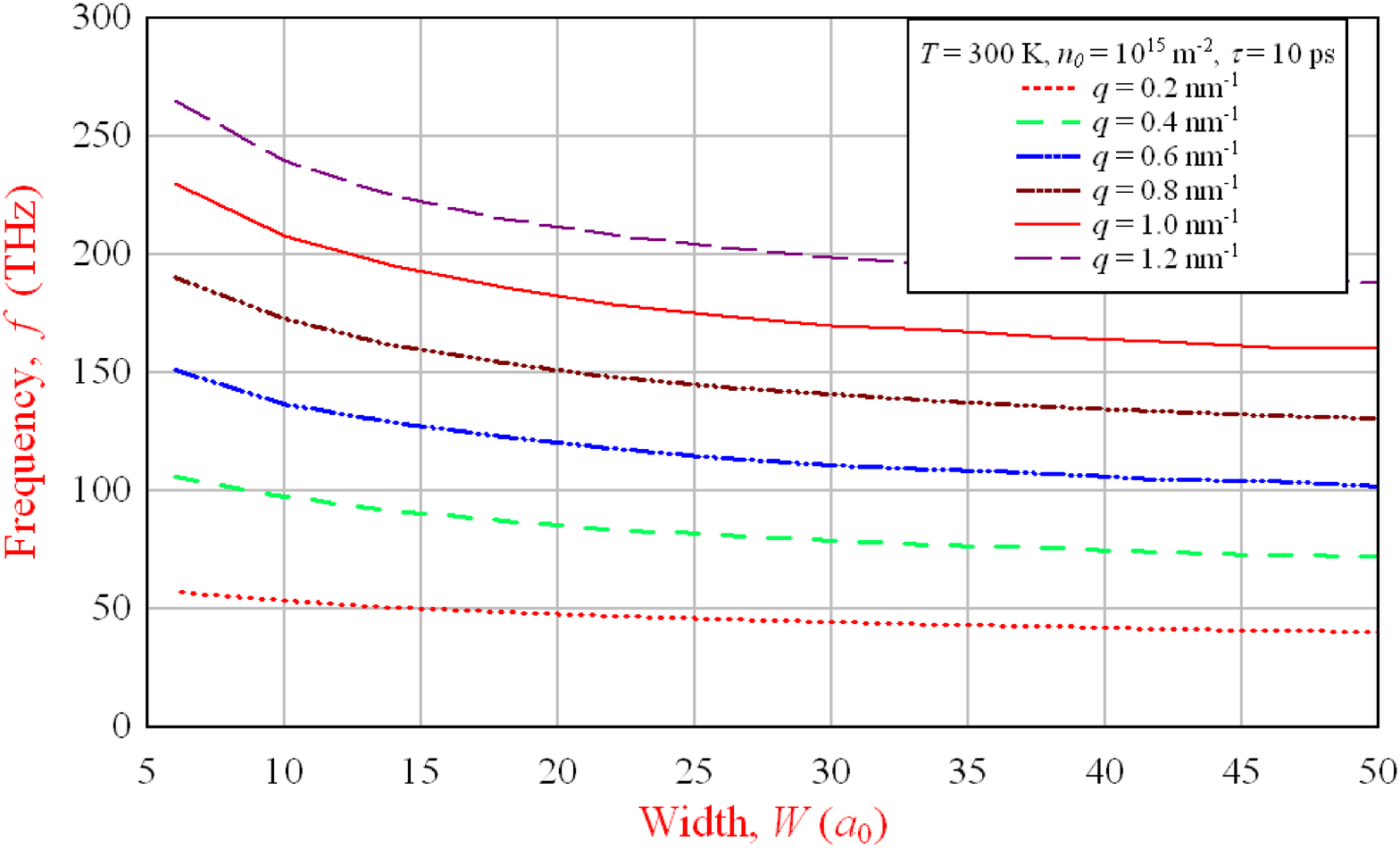}
\caption{The dependence of the plasmon frequency $f$ on the width of
the nanoribbon $W$, for the different wave vectors $q_{x}$ at the
fixed dissipation time $\tau$ corresponding to the damping,
temperature $T$ and doping electron density $n_{0}$.} \label{Fig5}
\end{figure}

 The dependence of the critical density of the QDs $N_{c}$ required for
the amplification of the signal on the wave vector $q_{x}$ for the
different doping electron densities $n_{0}$ at the fixed width of
the nanoribbon, temperature and dissipation time $\tau $
corresponding to the damping, obtained using Eq.~(\ref{N02}) is
presented in Fig.~\ref{Fig1}. According to Fig.~\ref{Fig1}, $N_{c}$
decreases as $q_{x}$ and $n_{0}$ increase. Let us mention that at
$q_{x}$ larger than $0.6\ \mathrm{nm^{-1}}$ there is almost no
difference between the values of $N_{c}$ corresponding to the
different doping electron densities $n_{0}$, and for large $q_{x}$
$N_{c} $ converges to approximately $18\ \mathrm{\mu m^{-2}}$. In
Fig. \ref{Fig2} the dependence of the critical density of the QDs
$N_{c}$ required for the amplification of the signal on the wave
vector $q_{x}$ for the different dissipation time corresponding to
the damping at the fixed width of the nanoribbon, temperature and
doping electron densities obtained using
Eq.~(\ref{N02})   is shown. As it follows from Fig.~\ref{Fig2}, $N_{c}$ decreases as $%
q_{x}$ and $\tau $ increase. This means that higher damping
corresponds to higher $N_{c}$. According to Fig.~\ref{Fig2},
starting with $q_{x}\approx 1.0\ \mathrm{(nm)^{-1}}$, $N_{c}$
depends very weakly on $q_{x}$, converging to some constant values
that depend on the value of $\tau $. The dependence of the critical
density of the QDs $N_{c}$ required for the amplification of the
signal on the width of the nanoribbon $W$ at the different wave
vector  for the fixed dissipation time  corresponding to the
damping,
temperature  and doping electron densities  obtained using Eq.~(%
\ref{N02}) is  displayed in Fig.~\ref{Fig3}. From Fig.~\ref{Fig3} we
can conclude that $N_{c}$ increases as $W$ increases and decreases
as $q_{x}$
increases. When $W$ increases, the values of $N_{c}$ stronger depend on $%
q_{x}$. The dependence of the critical density of the QDs $N_{c}$
required for the amplification of the signal  on the frequency $f$
at the different dissipation time  corresponding to the damping for
the fixed temperature  and doping electron density obtained using
Eq.~(\ref{N02}) is shown in Fig.~\ref{Fig4}. As it is demonstrated in Fig.~%
\ref{Fig4}, $N_{c}$ increases as $f$ and $\tau $ decrease. According
to Fig.~\ref{Fig4}, starting with $f \approx 140\ \mathrm{THz}$,
$N_{c}$ depends very weakly on frequency and converges to some
constant values that depend on the value of $\tau $. The dependence
of the plasmon frequency $f$ on the width of the nanoribbon $W$, for
the different wave vectors  at the fixed dissipation time
corresponding to the damping,
temperature and doping electron density  obtained using Eq.~(\ref%
{pl}) is presented in Fig.~\ref{Fig5}. According to Fig.~\ref{Fig5},
the plasmon frequency $f$ increases as $q_{x}$ increases and the
width of the nanoribbon $W$ decreases. If in Eq.~(\ref{N02}), the
imaginary part of the dielectric function would not depend on the
width $W$, $N_{c}$ would  not depend on
 $W$. However, due to the complicated dependence of $\mathrm{%
Im}\varepsilon (\omega ,q_{x})$ on $W$ through $V_{0,0}(q_{x})$
given by
Eq.~(\ref{V002}), this dependence  exists. For the damping time we use $%
\tau =5\ \mathrm{ns}$, $\tau =10\ \mathrm{ns}$, and $\tau =20\
\mathrm{ns}$,
because such damping for graphene was obtained in the experimental studies~%
\cite{Neugebauer,Orlita,Dubinov,Emani}. One can conclude from Figs.~\ref%
{Fig2} and~\ref{Fig4}, that $N_{c}$ decreases when the damping time
$\tau $ increases.

Let us mention that we used the parameters for PbS and PbSe QDs to
calculate $N_{c}$, because among different materials for the QDs the
PbS and PbSe QDs demonstrate the lowest transition
frequency~\cite{Auxier}, which can be in the resonance with the
plasmon in graphene nanoribbon in the IR region of spectrum.
According to Ref.~\onlinecite{Plum}, the transition frequency for
the QDs depends on the radius of the QDs. The PbS QDs with the radii
$2 \ \mathrm{nm}$---$5 \ \mathrm{nm}$ have the transition
frequencies $231$ and $194 \ \mathrm{THz}$~\cite{Plum}. For our
calculations we use PbS and PbSe QDs synthesized with the radii up
to $8 \ \mathrm{nm}$, which can provide the transition frequency $f
\approx 170 \ \mathrm{THz}$~\cite{Stockman_QD}. Let us mention that
changing the radius of the QDs, we can change the frequency of the
QDs resonant to the plasmon frequency in graphene nanoribbon
controlled by the wave vector $q_{x}$, and, therefore, we can
control $N_{c}$ by the radius of the $QDs$. The density of PbS QDs
with the diameter $3.2 \ \mathrm{nm}$ applied for the amplification
of plasmons in a gold film in the experiment~\cite{Plum} was $4\times 10^{6} \ \mathrm{%
\mu m^{-2}}$. According to Figs.~\ref{Fig1}-\ref{Fig4}, in the
graphene nanoribbon-based spaser there are possibilities to achieve
much less densities of the PbS QDs necessary for amplification than
in the gold film-based spaser.

Let us mention that in our calculations we take into account the
temporal and spatial dispersion of the dielectric function of
graphene nanoribbon in the random phase approximation
\cite{Brey_Fertig_01,Brey_Fertig}. The effects of spatial dispersion
are very important for the properties of spaser based on a flat
metal nanofilm~\cite{Larkin}. Taking into account the spatial
dispersion of the dielectric function of a metal surface in the
local random phase approximation allows to conclude that the strong
interaction of QD with unscreened metal electrons in the surface
nanolayer causes enhanced relaxation due to surface plasmon
excitation and Landau damping in a spaser based on a flat metal
nanofilm~\cite{Larkin}. And we assume that taking into account the
spatial dispersion of the dielectric function of graphene nanoribbon
is also very important to calculate the minimal population inversion
needed for the net SP amplification in the graphene nanoribbon based
spaser.

The advantages of the graphene nanoribbon based spaser are wide
frequency generation region from THz up to IR, small damping --- low
threshold for pumping, possibility of control by the gate. While we
perform our
calculations for IR radiation corresponding to the transition frequencies $%
170\ \mathrm{THz}$, the graphene-based spaser can work at the frequencies
much below that this one including THz regime.


\acknowledgments

The authors are grateful to M.~I. Stockman for valuable discussions.
The work was supported by PSC CUNY under Grant No.  65572-00 43.


\appendix


\section{The dielectric function of graphene nanoribbon}

\label{ap.diel}

For the armchair graphene nanoribbon the dielectric function $\varepsilon
(q_{x},\omega ,\beta ,\mu )$ in the one-band approximation in the random
phase approximation is given by~\cite{Brey_Fertig}
\begin{eqnarray}  \label{epsilon}
\varepsilon _{00}(q_{x},\omega ,\beta ,\mu _{g})=1-V_{0,0}(q_{x})\Pi
_{0,0}(q_{x},\omega ,\beta ,\mu _{g})\ ,
\end{eqnarray}
where $V_{0,0}(q_{x})$ is the Coulomb matrix element, and  the
polarizability $\Pi _{0,0}(q_{x},\omega )$ can be approximated by
\begin{eqnarray}  \label{pi}
\Pi _{0,0}(q_{x},\omega ,\beta ,\mu _{g})=-\frac{g_{s}}{\pi
}\frac{\hbar
v_{F}q_{x}}{\hbar ^{2}(\omega (\omega +i\gamma )-(v_{F}q_{x})^{2})}%
f_{1}(q_{x},\beta ,\mu _{g})
\end{eqnarray}
with
\begin{eqnarray}  \label{f1}
f_{1}(q_{x},\beta ,\mu _{g})=\frac{1}{\hbar \beta v_{F}}\left(
-\beta \hbar v_{F}q_{x}+2\ln \left( \frac{1+e^{-\beta \mu
_{g}}}{1+e^{-\beta (\hbar v_{F}q_{x}+\mu _{g})}}\right) \right) \ ,
\end{eqnarray}
where $g_{s}=2$ is the spin degeneracy factor, $\beta =1/(k_{B}T)$, $k_{B}$
is the Boltzmann constant and $\mu _{g}$ is the chemical potential
controlled by the doping. The chemical potential can be calculated as $\mu
=(\pi n_{0})^{1/2}\hbar v_{F}$, where electron concentration is given by $%
n_{0}$ and $v_{F}=\sqrt{3}a_{0}t/(2\hbar )\approx 10^{8}\
\mathrm{cm/s}$ that is
the Fermi velocity of electrons in graphene, where $a_{0} =2.46\ \mathrm{%
\mathring{A}}$ is a lattice constant and the value of the overlap integral
between the nearest carbon atoms is $t\approx 2.71\ \mathrm{eV}$ \ \cite%
{Lukose}.

Let us mention that at $T=0 \ \mathrm{K}$ we have
$f_{1}(q,\beta,\mu_{g}) = -q$. The temperature $T=300\ \mathrm{K}$
was used in Ref.~\onlinecite{Brey_Fertig}.

The Coulomb matrix element $V_{0,0}(q_{x})$ is given by~\cite{Brey_Fertig}
\begin{eqnarray}  \label{V00}
V_{0,0}(q_{x}) = \int_{0}^{1}du \int_{0}^{1}du^{\prime}
V(q_{x}W|u-u^{\prime}|) \ ,
\end{eqnarray}
where $W$ is the width of graphene nanoribbon in the $y$ direction.
The one-dimensional Fourier transform of the Coulomb interaction has
the form~\cite{Li_DasSarma}
\begin{eqnarray}  \label{V001}
V(q_{x}|y-y^{\prime}|) = \frac{2ke^{2}}{\varepsilon_{g}}K_{0}(q_{x}|y-y^{%
\prime}|) \ ,
\end{eqnarray}
where $e$ is the charge of an electron, $\varepsilon_{g} = 2.5$ is
the dielectric constant of graphene, $K_{0}(y)$ is the zeroth-order
modified Bessel function of the second kind, which diverges as $-\ln
y $ when $y$ goes to zero.

Using the definition for $V(q_{x}W|u-u^{\prime}|)$ from
Eq.~(\ref{V001}), we obtain from Eq.~(\ref{V00}):
\begin{eqnarray}  \label{V002}
V_{0,0}(q_{x}) = \frac{2ke^{2}}{\varepsilon_{g}} \int_{0}^{1}du
\int_{0}^{1}du^{\prime} K_{0}(q_{x}W|u-u^{\prime}|) \ .
\end{eqnarray}

We are interested to obtain the dynamical dielectric function
$\varepsilon (q_{x},\omega,\beta,\mu)$ at the  frequencies $\omega
\gg v_{F}k$.  After
the substitution of $\Pi_{0,0}(q_{x},\omega,\beta,\mu_{g})$ from Eq.~(\ref%
{pi}) and $V_{0,0}(q_{x})$ from Eq.~(\ref{V002}) into Eq.~(\ref{epsilon}),
and taking into account Eq.~(\ref{f1}) one can obtain the dielectric
constant for the graphene nanoribbon $\varepsilon_{00}(q_{x},\omega,\beta,%
\mu_{g}) \equiv \varepsilon (q_{x},\omega)$.



\section{The electric field of a plasmon in a graphene nanoribbon}

\label{ap.el}

$\vec{E}(x,y,z)$ can be found from the equation:
\begin{eqnarray}  \label{eqD}
\nabla\cdot \vec{D}(x,y,z) = \frac{\partial D_{x}(x,y,z)}{\partial x} +
\frac{\partial D_{y}(x,y,z)}{\partial y} + \frac{\partial D_{z}(x,y,z)}{%
\partial z} = 0 \ .
\end{eqnarray}
If there are two materials contacting each other along the plane there are
the boundary conditions at the plane of the contact: $D_{n1} = D_{n2}$ and $%
E_{t1} = E_{t2}$, where $D_{n1}$ and $D_{n2}$ are the the normal to the
contact plane components of $\vec{D}$, and $E_{t1}$ and $E_{t2}$ are tangent
to the contact plane components of $\vec{E}$.

Using the notation $u=x-x^{\prime}$, $v=y-y^{\prime}$, we have
\begin{equation}
\vec{D}(x,y,z) = \cases{ \int_{-\infty}^{+\infty} du \int_{-W/2}^{+W/2} dv
\varepsilon (u,v) \vec{E}(x-u,y -v,z) & at $z=0$ and $-W/2<y<W/2$ \cr
\varepsilon_{d} \vec{E}(x,y,z) & at $z<0$ or $z>0$ or $y<-W/2$ or $y>W/2$
\cr } \ .  \label{gef}
\end{equation}

At $z=0$ and $-W/2<y<W/2$ we have from Eqs.~(\ref{eqD}) and~(\ref{gef})
\begin{eqnarray}  \label{divD2}
\int_{-\infty}^{+\infty} du \int_{-W/2}^{+W/2} dv \varepsilon(u,v) \left(%
\frac{\partial E_{x}(x-u,y-v,z)}{\partial x} + \frac{\partial
E_{y}(x-u,y-v,z)}{\partial y} +\frac{\partial E_{z}(x-u,y-v,z)}{\partial z}%
\right) = 0 \ ,
\end{eqnarray}
which can be presented as
\begin{eqnarray}  \label{divD4}
\int_{-\infty}^{+\infty} du \int_{-W/2}^{+W/2} dv \varepsilon(u,v) \mathrm{%
div} \vec{E}(x-u,y-v,z) = 0 \ ,
\end{eqnarray}
If we do the Fourier transformation of Eq.~(\ref{divD4}) and use the
property of the Fourier image of the convolution, we obtain
\begin{eqnarray}  \label{divE1}
\nabla \vec{E}(x,y,z) = 0 \ .
\end{eqnarray}
Let us mention that for $z<0$ or $z>0$ or $y<-W/2$ or $y>W/2$ it is obvious
that Eq.~(\ref{divE1}) follows from Eqs.~(\ref{eqD}) and~(\ref{gef}).

If we define the potential $\vec{E}(x,y,z) = - \nabla \varphi (x,y,z)$, we
obtain from Eq.~(\ref{divE1}) for $z<0$ and $z>0$
\begin{eqnarray}  \label{lap}
\Delta \varphi (x,y,z) = 0 \ .
\end{eqnarray}
The solution of Eq.~(\ref{lap}) for $z\neq 0$ will be written as
\begin{eqnarray}  \label{lap1}
\varphi (x,y,z) = \psi (k_{x}x +k_{y}y \pm i\sqrt{k_{x}^{2} +k_{y}^{2}}|z|)
= \psi (v) \ ,
\end{eqnarray}
where $v = k_{x}x +k_{y}y \pm i\sqrt{k_{x}^{2} +k_{y}^{2}}|z|$. Then
we have for $\vec{E}(x,y,z)$ where $\psi^{\prime} = d\psi(w)/dw$
\begin{eqnarray}  \label{lap3}
\vec{E}(x,y,z) = -\psi^{\prime}(k_{x}x +k_{y}y \pm i\sqrt{k_{x}^{2}
+k_{y}^{2}}|z|) \vec{b}_{0} \ ,
\end{eqnarray}
where $ \vec{b}_{0} = (k_{x},k_{y},\pm i
\mathrm{sign}(z)\sqrt{k_{x}^{2} +k_{y}^{2}})$.
From the boundary conditions, we have $k_{x} = iq_{x}$, $k_{y} = \pm iq_{y}$%
, where for the armchair nanoribbon we have from
Ref.~\onlinecite{Brey_Fertig_01} $q_{yn} =
2\pi/(3a_{0})\left((2M+1+n)/(2M+1)\right)$ at the width $W= (3M+1)
a_{0}$, where $a_{0}$ is the graphene lattice constant defined
above, $M$ is the integer. We will use $n=1$.

Defining $\alpha = \sqrt{q_{x}^{2} + q_{y}^{2}}$ and using $\psi (w)
= E_{0}e^{w}/2$, we obtain for $\vec{E}(x,y,z)$ from
Eq.~(\ref{lap3}):
\begin{eqnarray}  \label{E}
\vec{E}(x,y,z) = -\frac{E_{0}}{2}e^{iq_{x}x-\alpha
|z|}\left(e^{iq_{y}y} \vec{b}_{1} + e^{-iq_{y}y} \vec{b}_{2} \right)
\ ,
\end{eqnarray}
where $\vec{b}_{1} = (iq_{x},iq_{y},-\alpha \mathrm{sign}(z))$ and
$\vec{b}_{2} = (iq_{x},-iq_{y},-\alpha \mathrm{sign}(z))$.

\end{document}